\documentclass[preprint,11 pt,times]{elsarticle}
\usepackage[margin=2cm]{geometry}
\usepackage{amsmath}
\usepackage{amsfonts}
\usepackage{amssymb}
\usepackage{graphicx}
\usepackage{lmodern}
\usepackage{hyperref}
\usepackage{dcolumn}
\usepackage{bm}
\usepackage[mathlines]{lineno}
\usepackage{float}
\usepackage[T1]{fontenc}
\usepackage{subfigure}
\usepackage{natbib}
\journal{Journal of physics G}

\usepackage{numcompress}

\DeclareUnicodeCharacter{2212}{\ensuremath{-}}

\begin{document}
\newcolumntype{P}[1]{>{\centering\arraybackslash}p{#1}}
\title{\textbf{Revisiting the Dirac Nature of Neutrinos in the Light of $\Delta(27)$ and Cyclic Symmetries.}}



\author[mymainaddress]{Manash Dey}
\ead{manashdey@gauhati.ac.in}

\author[mymainaddress]{Subhankar Roy\corref{mycorrespondingauthor}}
\cortext[mycorrespondingauthor]{Corresponding author}
\ead{subhankar@gauhati.ac.in}

\address[mymainaddress]{Department of Physics, Gauhati University, India, 781014}

\begin{abstract}
Amid the uncertainty regarding the fundamental nature of neutrinos, we adhere to the Dirac description, and construct a model in the framework of $\Delta(27)$ symmetry. The model successfully accounts for the hierarchical patterns of both charged lepton and neutrino masses. The neutrino mass matrix exhibits four texture zeroes, and the associated mixing scheme aligns with the experimental data, notably controlled by a single parameter. 
\end{abstract}
\maketitle
\section{Introduction}
The neutrinos are an essential part of the Standard Model(SM) of particle physics. According to the SM, neutrinos are massless particles. However, a quantum mechanical phenomenon called Neutrino Oscillation, originally proposed by Bruno Pontecorvo in 1957\,\cite{Pontecorvo:1957cp}, suggests that the three mass eigenstates of neutrinos\,($\nu_{i= 1,2,3}$) mix together to generate the three flavour neutrino states\,($\nu_{l= e,\mu,\tau}$). In recent past, observations from various experiments\,\cite{ SNO:2002tuh, KamLAND:2002uet, Super-Kamiokande:1998uiq} confirmed the occurrence of neutrino oscillation, and it was seen that, for the oscillation to occur, the neutrinos should have non zero and non-degenerate masses. Hence, the SM has two primary shortcomings: it cannot fully explain the reason why neutrinos of different flavours mix together peculiarly, and it cannot account for the extremely small masses of neutrinos. The search for answers to these questions has taken us to the realm of neutrino mass matrix textures. The term ``neutrino mass matrix'' refers to a mathematical structure which contains information of the neutrino masses and mixing, and the term ``neutrino mass matrix texture'' refers to certain patterns among the elements of the neutrino mass matrix. These patterns serve to reduce the number of free parameters in the mass matrix and help to draw important correlations among the observable quantities. To understand the neutrino mass matrix textures, researchers have primarily investigated the idea that neutrinos are their own antiparticles, known as Majorana neutrinos. This assumption is motivated by the fact that, for the Majorana nature, the equations describing neutrino masses\,($\sim \bar{\nu_L} \nu_R + h.c$) become simpler because they can be constructed solely from left-handed neutrinos\,($\nu_L$), represented as $\sim \bar{\nu_L} \nu^c_L + h.c$. Majorana nature allows us to equate $\nu_R$\,(right-handed neutrinos) with $\nu^c_L$\,(charge conjugate of the left-handed neutrino), eliminating the need for an additional right-handed component. The Majorana mass term\,($\sim \bar{\nu_L} \nu^c_L + h.c$), is a lepton number violating term, it violates the lepton number by two units. In the SM, this term applies only to neutrinos, as they are electrically neutral. However, it is to be emphasized that the Majorana nature of neutrinos is subject to experimental verification. In this light, experiments on neutrino less double-beta\,($0\nu\beta\beta$) decay are considered to be the most promising ones\,(for a detailed review, refer to\,\cite{Agostini:2022zub,Barabash:2023dwc, Gómez-Cadenas2023}). Although attempts to identify $0\nu\beta\beta$ decay have been going on for almost a century, there is, so far no evidence to support it\,\cite{Gómez-Cadenas2023}.
This leaves ample room for the discussions related to the Dirac nature. In the present work, we adhere to the Dirac nature of the neutrinos.

	 The plan of our work is as follows: In Section \ref{framework}, we discuss the framework on which the model is built. In addition, we also discuss the standard parametrization of the lepton mixing matrix predicted from our model. In Section \ref{NA}, we shed light on the numerical analysis and findings of our study. Finally, in Section \ref{Summary}, we highlight the summary and discussions of the present work.
\section{Framework \label{framework}}
We extend the field content of the SM with right handed neutrinos\,($\nu_{R}$), and four additional complex scalar fields, namely, $\psi, \phi$, $\kappa$ and $\sigma$. Our model is grounded in the framework of $\Delta(27)$ symmetry. The discrete symmetry group $\Delta(27)$\,\cite{Branco:1983tn, Luhn:2007uq, Ma:2007wu, Ishimori:2010au, Abbas:2014ewa, Chen:2015jta, CentellesChulia:2016fxr, Vien:2020hzy,Dey:2023bfa} has 11 irreducible representations, namely, one triplet\,($3$), one anti-triplet \,($\bar{3}$) and nine singlet\,($1_{p,q}$) representations, where $p$ and $q$ are integers, and can take values from 0 to 2. We further enrich the model with $Z_N$\,\cite{AristizabalSierra:2014irc, CentellesChulia:2016fxr, CarcamoHernandez:2020udg} symmetries, specifically $Z_3$\,\cite{Ma:2004yx, Hu:2006wk}, $Z_7$ and $Z_{10}$\,\cite{CarcamoHernandez:2020udg, Dey:2023rht}, whose significance will be discussed in this section. For reviews on discrete flavour symmetries and their implications in neutrino masses and mixing, see Refs.\cite{King:2013eh, King:2017guk, Feruglio:2019ybq}, and for a recent review, see Ref.\cite{Chauhan:2023faf}. A brief description of the product rules associated with $\Delta(27)$  and $Z_N$ symmetries are shown in the Appendix \ref{appA}. 
\subsection{Model \label{model}}
We tabulate the transformation properties of all the field contents in Table\,\ref{table:1}.
\begin{table}[h!]
\centering
\begin{tabular}{p{1.3cm}p{0.7cm}p{2.5cm}p{1.2cm}p{1cm}p{1cm}p{1cm}p{1cm}p{0.7cm}}
\hline
\noalign{\vskip 0.55mm}
Fields & $\bar{D}_{l_{L}}$ & \quad\quad$l_{R}$ & $H$ &$\nu_R$ &$\psi$ &  $\phi$ &$\kappa$&$\sigma$\\
\hline
\noalign{\vskip 1mm}
$\Delta(27)$ & 3 & $(1_{00},1_{22},1_{11})$ & $1_{00}$ &3& $\bar{3}$ & 3 &3&$1_{00}$ \\
\hline
$Z_{3}$ & $\omega^*$ & \,(1,\quad 1,\quad $\omega^*$) & 1 &$\omega$& $\omega$& 1 &$\omega^*$&$\omega$ \\
\hline
$Z_{7}$ & 4 &\quad\quad\, 3 & 0 & 3 & 0 & 0 & 0 & 0 \\
\hline
$Z_{10}$ & 1 &\,(9,\quad 2,\quad 4) & 0 &6& 4& 3 &2&1 \\
\hline
$SU(2)_{L}$ & 2 &\quad\quad\, 1 & 2 &1& 1& 1 &1&1 \\
\hline
\end{tabular}
\caption{ The transformation properties of the field contents under $\Delta(27)\times Z_{3} \times Z_7 \times Z_{10} \times SU(2)_L$} 
\label{table:1}
\end{table}
The Yukawa Lagrangian invariant under $\Delta(27)\times Z_{3} \times Z_7 \times Z_{10} \times SU(2)_L$ symmetry is presented below,
\begin{eqnarray}
- \mathcal{L}_{Y} &=& \frac{y_e}{\Lambda^7} (\bar{D}_{l_L}\psi)\,H\, e_{R}\,\sigma^6+\,\frac{y_{\mu}}{\Lambda^4} (\bar{D}_{l_L}\psi)\,H\,\mu_{R}\,\sigma^3+\,\frac{y_{\tau}}{\Lambda^2} (\bar{D}_{l_L}\psi)\,H \,\tau_{R}\,\sigma+ \frac{y_s}{\Lambda}\,(\bar{D}_{l_L}\phi)\,\tilde{H}\,\nu_{R}+\frac{y_a}{\Lambda}\,(\bar{D}_{l_L}\phi)\nonumber\\&&\,\tilde{H}\,\nu_{R}+\frac{y_d}{\Lambda^2}\,(\bar{D}_{l_L}\nu_{R})\,\tilde{H}\,\kappa\,\sigma, \nonumber
\label{Lagrangian}
\end{eqnarray}
where, $\Lambda$ is the cut-off scale of the theory.

	The cyclic symmetries $Z_{10}$ and $Z_3$ enforce coupling of the complex scalar field $\psi$ to the charged leptons, and fields $\phi$ and $\kappa$ to the right-handed neutrinos. They also constrain unfavorable terms and next-to-leading order (NLO) corrections in the Yukawa Lagrangian ($\mathcal{L}_{Y}$). The $Z_7$ symmetry forbids possible Majorana mass terms from appearing in the model.

	It is important to mention that the symmetries, $Z_{10}$ and $Z_{3}$, are broken by the vacuum expectation value\,(VEV) of the scalar singlet $\sigma$. This breaking of symmetry is crucial for the explanation of the charged lepton mass hierarchy \,\cite{Froggatt:1978nt,Babu:1999me,Lykken:2008bw,Hagedorn:2010mq,Ganguly:2022cbo,Bonilla:2023wok,CarcamoHernandez:2024vcr}.
	 
	 We assume that the scalar fields develop the VEVs in the following directions:  $\langle \psi \rangle = v_{\psi}(1,1,1)$, $\langle \phi \rangle = v_{\phi}(0,1,0)$, $\langle \kappa \rangle = v_{\kappa}(1,1,1)$,  $\langle H \rangle = v_h$ and  $\langle \sigma \rangle = v_{\sigma}$. A detailed discussion on the scalar potential for these alignments is provided in Appendix \ref{appB}, where it is shown that, starting from the minimization conditions, one can derive the following equations involving the VEVs ($v_\psi, v_\phi, v_\kappa, v_h $ and $v_\sigma$ ) and additional twenty-five free parameters, expressed in terms of $\mu$'s and $\lambda$'s.
	 \begin{align}
\label{mu1}
\mu_H^2&= 2 \lambda^H_1 v_h^2+3 \lambda ^{H \kappa} v_{\kappa}^2+\lambda ^{H \sigma} v_{\sigma}^2+3 \lambda ^{H \psi} v_{\psi}^2+\lambda ^{H \phi} v_{\phi}^2,\\
\mu^2_{\psi}&=6( \lambda _1^{\psi }+ \lambda _3^{\psi })v^2_{\psi}+v_h^2 \lambda ^{H\psi}+3 v_{\kappa}^2( \lambda _1^{\psi \kappa }+ \lambda _3^{\psi \kappa }+\lambda _7^{\psi \kappa })+\lambda _1^{\psi \phi } v_{\phi}^2+\lambda ^{\psi \sigma } v_{\sigma}^2,\\
\mu _{\phi }^2&=2(\lambda _1^{\phi }+\lambda _2^{\phi })v^2_{\phi}+v_h^2 \lambda ^{H\phi}+3 v_{\kappa}^2 \lambda _1^{\phi \kappa }+3 \lambda _1^{\psi \phi } v_{\psi}^2+\lambda ^{\phi \sigma } v_{\sigma}^2,\\
\mu _{\kappa }^2&=6(\lambda _1^{\kappa }+\lambda _3^{\kappa })v^2_{\kappa}+v_h^2 \lambda ^{H\kappa}+\lambda ^{\kappa \sigma} v_{\sigma}^2+\lambda _1^{\phi \kappa } v_{\phi}^2+3 v_{\psi}^2(\lambda _1^{\psi \kappa } +\lambda _3^{\psi \kappa }+ \lambda _7^{\psi \kappa }),\\
\mu_{\sigma}^2&=2 \lambda^{\sigma}_1 v^2_{\sigma}+v_h^2 \lambda ^{H \sigma}+3 v_{\kappa}^2 \lambda^{\kappa \sigma}+\lambda^{\phi \sigma } v_{\phi}^2+3 \lambda ^{\psi \sigma } v_{\psi}^2.
\label{mu5}
\end{align}
The above equations are useful in obtaining the desired values of the VEVs. The specified directions of the VEVs are crucial for generating the correct mixing pattern, as they ensure that the charged lepton mass matrix\,($M_L$) as non diagonal. The form of $M_L$ is shown below,
\begin{equation}
 M_{L} =  v_h v_{\psi}
 \begin{bmatrix}
\frac{ y_e v_{\sigma}^6}{\Lambda^7} & \frac{ y_{\mu}v_{\sigma}^3}{\Lambda^4} & \frac{y_{\tau}v_{\sigma}}{\Lambda^2} \\
\frac{ y_e v_{\sigma}^6}{\Lambda^7}& \frac{ \omega^2\, y_{\mu} v_{\sigma}^3}{\Lambda^4} &  \frac{\omega\, y_{\tau} v_{\sigma}}{\Lambda^2}\\
 \frac{ y_e v_{\sigma}^6}{\Lambda^7}& \frac{\omega \,y_{\mu} v_{\sigma}^3}{\Lambda^4}& \frac{\omega^2\, y_{\tau} v_{\sigma}}{\Lambda^2} \\
 \end{bmatrix},
 \label{clmatrix}
\end{equation}
where, 
\begin{equation}
\omega= e^{2\pi i/3}\nonumber.
\end{equation}
To have the information about the charged lepton masses, we diagonalise $M_L$ in the following manner,
\begin{equation}
U_L^{\dagger} M_L U_R=
 \begin{bmatrix}
 \frac{\sqrt{3} v_h v_{\psi} v_{\sigma}^6  y_e}{\Lambda^7} & 0 &  0   \\
0& \frac{\sqrt{3} v_h v_{\psi} v_{\sigma}^3  y_{\mu}}{\Lambda^4}&0\\
 0& 0 & \frac{\sqrt{3} v_h v_{\psi} v_{\sigma}  y_{\tau}}{\Lambda^2}  \\
 \end{bmatrix},
 \label{cldmatrix}
\end{equation}
where, $U_L$ and $U_R$ are left handed  and right handed charged lepton diagonalising matrices respectively. The form of $U_L$ and $U_R$ is shown below,
\begin{equation}
 U_L=\frac{-1}{\sqrt{3}}
 \begin{bmatrix}
 1  & 1& 1  \\
1& \omega^2   & \omega \\
1& \omega  &\omega^2   \\
 \end{bmatrix},\quad U_R=\begin{bmatrix}
 -1  & 0& 0  \\
0& -1   & 0 \\
0& 0 &-1   \\
 \end{bmatrix}.
 \label{hmatrix}
 \end{equation}
From Eq.\,(\ref{cldmatrix}), it is observed that the charged lepton masses satisfy the following relation,
\begin{equation}
m_e \colon m_{\mu} \colon m_{\tau} \approx \lambda^5_m \, y_e \colon \lambda^2_m \,y_{\mu} \colon y_{\tau},
\label{clmh}
\end{equation}
where, $\lambda_m= v_{\sigma}/\Lambda$. The above relation is consistent with the observed hierarchy of the charged lepton masses.
\vspace*{0.4cm}

	In neutrino sector, the Dirac neutrino mass matrix($M_{\nu}$) takes the following form,
\begin{equation}
 M_{\nu} = 
 \begin{bmatrix}
\frac{y_d v_h v_k v_{\sigma}}{\Lambda^2} & 0 & \frac{v_h (y_s +y_a)v_{\phi}}{2\Lambda} \\
0& \frac{y_d v_h v_k v_{\sigma}}{\Lambda^2} &0\\
\frac{v_h (y_s -y_a)v_{\phi}}{2\Lambda}& 0 & \frac{y_d v_h v_k v_{\sigma}}{\Lambda^2} \\
 \end{bmatrix}.
 \label{nmatrix}
\end{equation}
The above mass matrix carries four texture zeroes\,\cite{Ahuja:2009jj,Fakay:2014nea,Singh:2018lao,Borgohain:2020csn}. The first study on Dirac neutrino mass matrices with texture zeroes was conducted in Ref.\cite{Hagedorn:2005kz}. In the literature, the Dirac neutrino mass matrix having four texture zeroes was investigated in earlier works, such as Ref.\cite{Memenga:2013vc}, which utilized the neutrinophilic two Higgs doublet model\,\cite{Ma:2000cc, Davidson:2009ha}, and Refs.\cite{Borah:2017dmk,Borah:2018nvu}, which employed the Dirac seesaw mechanism. Further investigations and models involving Dirac neutrinos can be found in Refs.\cite{Chang:2009mv, Chen:2012jg, Aranda:2013gga, Ding:2013eca, Holthausen:2013vba, Borah:2022enh, Zhao:2024lcg, Maharathy:2022gki, Berbig:2022hsm, Chen:2022bjb, Babu:2022ikf, CentellesChulia:2022vpz, Hazarika:2022tlc, Li:2022chc, Chowdhury:2022jde}. In the present work, our model successfully explains the charged lepton mass hierarchy, and we do not implement the Dirac seesaw mechanism for neutrino mass generation. While $A_4$\,\cite{Altarelli:2005yx, Altarelli:2010gt} symmetry is commonly employed for models involving the Dirac nature of neutrinos, we have opted for the rich framework of $\Delta(27)$ symmetry. The presence of the additional anti-triplet\,($\bar{3}$) representation in the $\Delta(27)$ symmetry, helps with a greater flexibility of tuning the neutrino mass matrix with a minimal field content. It is worth mentioning that in the literature\,\cite{Memenga:2013vc,Davidson:2009ha,Ma:2021bzl,Ma:2021szi}, the $U(1)$ group is widely employed to forbid the Majorana mass term. However, in our approach, we have opted for the $Z_7$ group instead. For the sake of simplicity, we redefine the elements of $M_{\nu}$ such that, $v_h (y_s -y_a)v_{\phi}/2\Lambda=r e^{i \gamma_1}, v_h (y_s +y_a)v_{\phi}/2\Lambda=r e^{i \gamma_2}$ and $y_d v_h v_k v_{\sigma}/\Lambda^2=r_3 e^{i \gamma_3}$. To understand the physics related to the neutrino mass matrix $M_{\nu}$, we define a hermitian matrix:
\begin{equation}
 H = M_{\nu} M^{\dagger}_{\nu}=
 \begin{bmatrix}
 r^2+r^2_3  & 0 & r r_3 e^{i(\gamma_3-\gamma_1)}+r r_3 e^{i(\gamma_2-\gamma_3)}  \\
0& r^2_3  &0\\
  r r_3 e^{-i(\gamma_3-\gamma_1)}+r r_3 e^{-i(\gamma_2-\gamma_3)}& 0 & r^2+r^2_3  \\
 \end{bmatrix},
 \label{hmatrix}
\end{equation}
which can be diagonalised with the help of a unitary matrix, $U_{\nu}$ in the following way: $U_{\nu}^{\dagger} H U_{\nu}= diag\,(m_1^2, m_2^2, m_3^2)$. Here, $U_{\nu}$ is the mixing matrix, and it takes the following form,
\begin{equation}
 U_{\nu}=\frac{1}{\sqrt{2}}
 \begin{bmatrix}
 - e^{i\rho} & 0 &  e^{i\rho}   \\
0& \sqrt{2} &0\\
 1& 0 & 1 \\
 \end{bmatrix},
 \label{mmatrix}
\end{equation}
where, 
\begin{equation}
\rho= (\gamma_2-\gamma_1)/2\nonumber.
\end{equation}

The three neutrino mass eigenvalues, $m_1, m_2$ and $m_3$ are obtained as: 
\begin{eqnarray}
m_1  &=& \sqrt{r^2 + r_3^2 - r\,r_3(2+2x)^{1/2}},\\
m_2 &=& r_3,\\
m_3 &=&\sqrt{r^2 + r_3^2 + r\,r_3(2+2x)^{1/2}},
\end{eqnarray}
where, $x=\cos\,(\gamma_1+\gamma_2-2 \gamma_3)$. It is noteworthy that the relation, $\Delta m^2_{31}=2 r r_3 \sqrt{2\,(1+x)}$, is always greater than zero, as $r>0$ and $r_3>0$. Hence, our model is true only for the normal ordering of neutrino masses. 
\vspace*{0.4cm}

The final lepton mixing matrix of our model is given by:
\begin{eqnarray}
\label{lpm}
U= U_L^{\dagger}.U_{\nu}=\left[
\begin{array}{ccc}
 \frac{-1+e^{i \rho }}{\sqrt{6}} & \frac{-1}{\sqrt{3}} & \frac{-1+e^{i \rho }}{\sqrt{6}} \\
 \frac{e^{i \rho }-\omega ^2}{\sqrt{6}} & \frac{-\omega }{\sqrt{3}} & -\frac{\omega ^2+e^{i \rho }}{\sqrt{6}} \\
 \frac{e^{i \rho }-\omega}{\sqrt{6}} & \frac{-\omega ^2}{\sqrt{3}} & -\frac{\omega +e^{i \rho }}{\sqrt{6}} \\
\end{array}
\right].
\end{eqnarray}
Needless to mention that in the above matrix, $\rho$ is the only free parameter. Before we try to fetch the information of the observational oscillation parameters from  $U$, we need to ensure that $U$ is represented as per standard or particle data group(PDG) parametrization\,\cite{ParticleDataGroup:2018ovx}. This is to be highlighted that most often, this important and necessary step is overlooked \,\cite{Memenga:2013vc, Roy:2014nua, Borah:2017dmk, Borah:2018nvu}. 

\subsection{Standard Parametrization \label{sp}}

A general $3\times 3$ unitary matrix carries nine free parameters and the lepton mixing matrix is no exception. However, the oscillation experiments deal only with four parameters out of nine. These are three mixing angles: solar angle ($\theta_{12}$), reactor angle\,($\theta_{13}$), and atmospheric angle\,($\theta_{23}$) as well as the  Dirac Charge Parity (CP) violating phase ($\delta$). The remaining parameters remain unobserved or unphysical. To ensure that the predictions are correct, we need to transform $U$ to $U_{std}$ following the transformation as shown below.
   
\begin{equation}
U_{std}=P^{*}_{\chi}.U.P^{*}_{\xi},
\label{stdmatrix}
\end{equation}

where, the matrices $P_{\chi}$ and $P_{\xi}$ carry the five unphysical parameters: $\chi_1$, $\chi_2$, $\chi_3$, $\xi_1$ and $\xi_2$, and these matrices are presented as: $P_{\chi}= diag(e^{i \chi_1}, e^{i \chi_2}, e^{i \chi_3})$ and $P_{\xi}= diag(e^{i \xi_1}, e^{i \xi_2}, 1)$. So, the $U_{std}$ appears as per PDG parametrization, as shown below,

\begin{eqnarray}
U_{std} &=&\left[
\begin{array}{ccc}
 c_{12} c_{13} & c_{13} s_{12} & e^{-i \delta } s_{13} \\
 -c_{23} s_{12}- e^{i \delta } c_{12} s_{13} s_{23} & c_{12} c_{23}-e^{i \delta } s_{12} s_{13} s_{23} & c_{13} s_{23} \\
 s_{12} s_{23}- e^{i \delta } c_{12} c_{23} s_{13} & -c_{12} s_{23} -e^{i \delta } c_{23} s_{12} s_{13} & c_{13} c_{23} \\
\end{array}
\right]
\end{eqnarray}
where, $s_{ij}=\sin\theta_{ij}$ and $c_{ij}=\cos\theta_{ij}$.

 After this transformation, the four observational parameters are predicted as shown in the following,

\begin{eqnarray}
\sin^2 {\theta_{12}}&=&\frac{1}{2-\cos(\rho)},\\
\sin^2 {\theta_{13}}&=&\frac{1+\cos(\rho)}{3},\\
\sin^2 {\theta_{23}}&=&\frac{\sqrt{3}\sin(\rho)+\cos(\rho)-2}{2\cos(\rho)-4},\\
\label{delta}
\delta &=&  \chi_1 (\rho)-\frac{\rho}{2}.
\end{eqnarray}
\emph{It is interesting to note the three mixing angles are dictated by a single parameter, $\rho$}, in contrast to\,\cite{Memenga:2013vc, Borah:2017dmk, Borah:2018nvu}. It seems that in the expression of $\delta$, in Eq.\,(\ref{delta}), $\chi_1$ is an independent parameter.  However, $\chi_1$ along with other four unphysical parameters can be worked out by solving the following five transcendental equations:  

\begin{eqnarray}
\label{10}
2 \sqrt{1-\cos (\rho )} \left(\sqrt{3} \sin \left(\xi _2+\chi _2\right)+3 \cos \left(\xi _2+\chi _2\right)\right)\nonumber\\-\sqrt{6} \left(\sin \left(\xi _1+\chi _2\right)-2 \sin \left(\rho -\xi _1-\chi _2\right)\right)\nonumber\\+3 \sqrt{2}\cos \left(\xi _1+\chi _2\right)=&0,\\
\label{11}
\sin \left(\xi _2+\chi _1\right)=&0,\\
\label{12}
\sin \left(\rho -\xi _1-\chi _1\right)+\sin \left(\xi _1+\chi _1\right)=&0,\\
\label{13}
3 \cos \left(\chi _2\right)-\sqrt{3} \left(2 \sin \left(\rho -\chi _2\right)+\sin \left(\chi _2\right)\right)=&0,\\
\label{15}
\sqrt{3} \left(2 \sin \left(\rho -\chi _3\right)+\sin \left(\chi _3\right)\right)+3 \cos \left(\chi _3\right)=&0,
\end{eqnarray}
which are obtained from the five standard parametrization conditions. This is to be noted that $\rho$ is the only free  parameter appearing in Eq.\,(\ref{10}) and Eqs.\,(\ref{12})-(\ref{15}). Hence, we understand that $\chi_1$ appearing in Eq.\,(\ref{delta}) is not independent, rather relies on $\rho$. \emph{Therefore, $\delta$ and the three mixing angles are basically guided by $\rho$ itself}. 

 	In the next section we will study the predictions of the present model for a wide range of data points.

\section{Numerical Analysis and Results \label{NA}}
	
\begin{figure}
  \centering
    \subfigure[]{\includegraphics[width=0.3
  \textwidth]{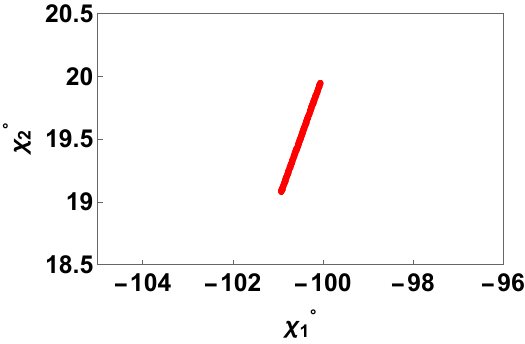}\label{fig:1(a)}} 
    \subfigure[]{\includegraphics[width=0.3
    \textwidth]{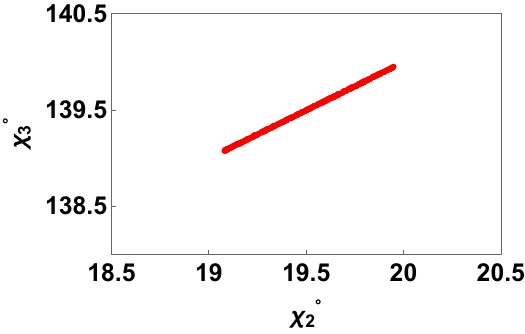}\label{fig:1(b)}} 
    \subfigure[]{\includegraphics[width=0.3
    \textwidth]{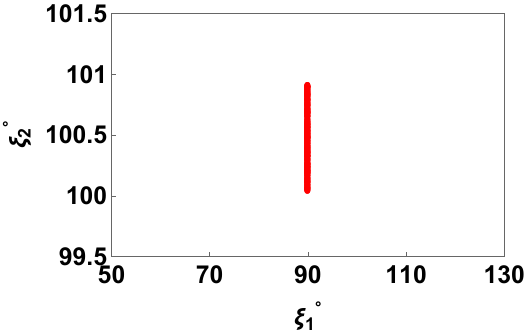}\label{fig:1(c)}} 
\caption{The plots between the unphysical phases\,(a) $\chi_1$ and $\chi_2$.\,(b) $\chi_2$ vs $\chi_3$.\,(c) $\xi_1$ and $\xi_2$.}
\label{fig:1}
\end{figure}

	 We numerically solve the Eqs.\,(\ref{10})\,to\,(\ref{15}) simultaneously in order to generate sufficiently large number of data points for the unphysical phases $\chi_1$, $\chi_2$, $\chi_3$, $\xi_1$ and $\xi_2$. The maximum and minimum values of the said phases and the parameter $\rho$ are tabulated in Table\,\ref{table:2}. We graphically represent these phases in Figures \ref{fig:1(a)} - \ref{fig:1(c)}. Though measurement of the unphysical phases is not the aim in the experiments, from the model building point of view, this information could be significant.
\begin{figure}
  \centering
    \subfigure[]{\includegraphics[width=0.31
  \textwidth]{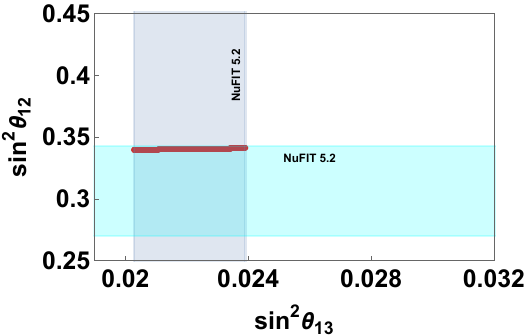}\label{fig:2(a)}} 
    \subfigure[]{\includegraphics[width=0.3\textwidth]{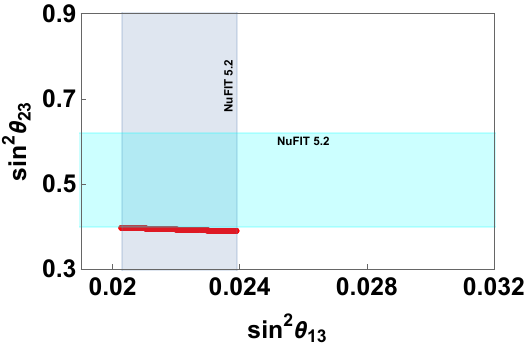}\label{fig:2(b)}} 
    \subfigure[]{\includegraphics[width=0.3\textwidth]{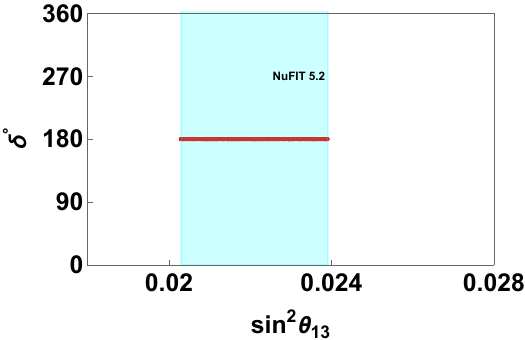}\label{fig:2(c)}}
    \subfigure[]{\includegraphics[width=0.32
    \textwidth]{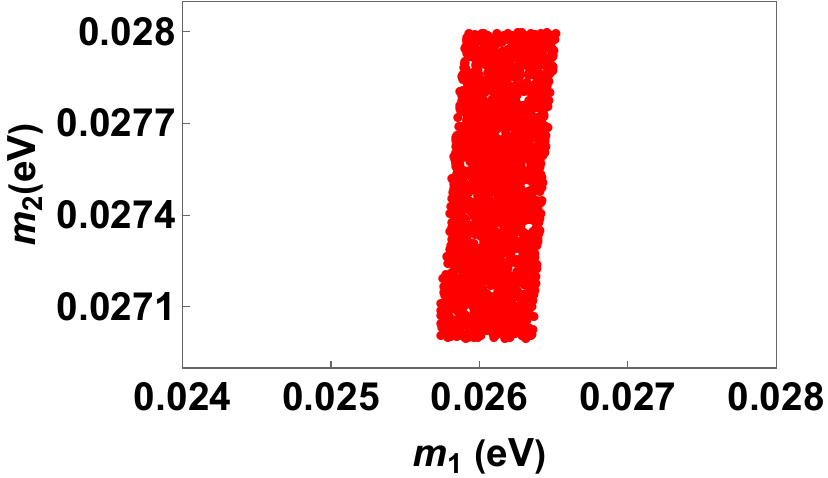}\label{fig:2(d)}}
    \subfigure[]{\includegraphics[width=0.32
    \textwidth]{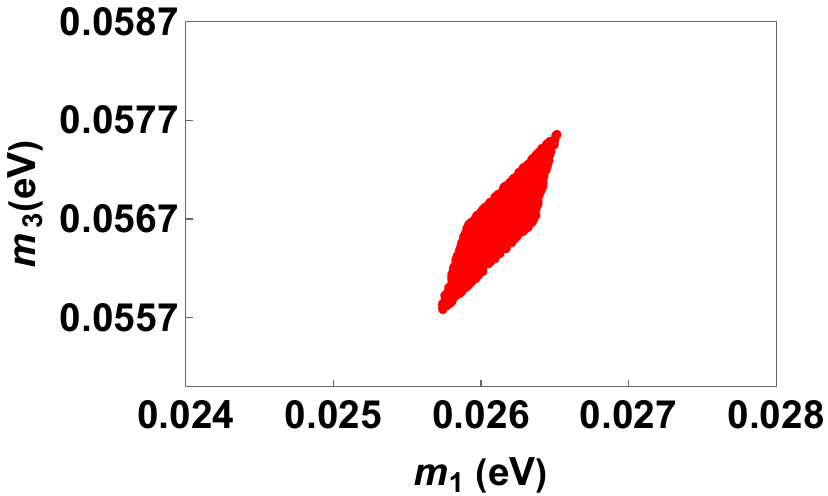}\label{fig:2(e)}}
    \subfigure[]{\includegraphics[width=0.345
    \textwidth]{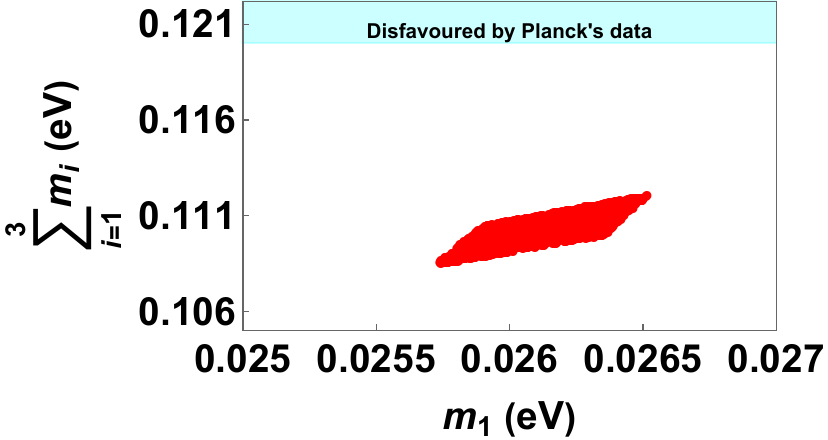}\label{fig:2(f)}}
     \subfigure[]{\includegraphics[width=0.285
  \textwidth]{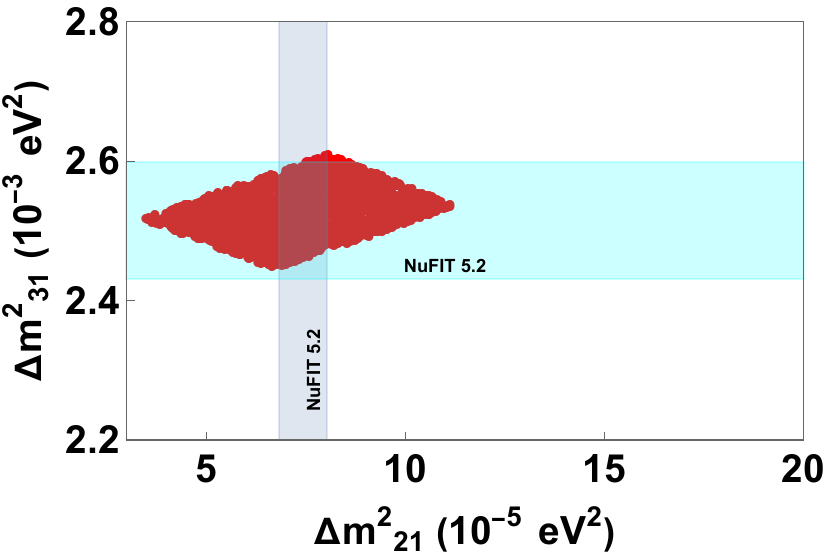}\label{fig:2(g)}}
\caption{The plots between the observable parameters:\,(a) $\sin^2 {\theta_{13}}$ and $\sin^2 {\theta_{12}}$.\,(b) $\sin^2 {\theta_{23}}$ and $\sin^2 {\theta_{13}}$.\,(c)  $\delta$ and $\sin^2 {\theta_{13}}$.\,(d) $m_2$ and $m_1$.\,(e) $m_3$ and $m_1$.\,(f) $\sum{m_i}$ and the lightest neutrino mass, $m_1$.\,(g) $\Delta m^2_{31}$ and $\Delta m^2_{21}$.}
\label{fig:2}
\end{figure}
\begin{table}[h]
\centering
\begin{tabular}{c|c|c}
\hline
Parameters  & Minimum Value & Maximum Value\\
\hline
$\rho$ &158.1730$^{\circ}$ & 159.9030$^{\circ}$\\
\hline
$\chi_1$ & -100.9130$^{\circ}$ & -100.0490$^{\circ}$ \\
\hline
$\chi_2$ & 19.0860$^{\circ}$ & 19.9513$^{\circ}$\\
\hline
$\chi_3$ & 139.0870$^{\circ}$ & 139.9510$^{\circ}$ \\
\hline
$\xi_1$ & 90$^{\circ}$ & 90$^{\circ}$\\
\hline
$\xi_2$ & 100.0490$^{\circ}$ & 100.9130$^{\circ}$\\
\hline
\end{tabular}
\caption{Represents the maximum and minimum values of the unphysical phases and the parameter $\rho$} 
\label{table:2}
\end{table}It is interesting to note that for the $3 \sigma$ range of $\sin^2{\theta_{13}}$, the observable $\sin^2{\theta_{12}}$ is predicted at the extreme end of the experimental $3 \sigma$ range\,\cite{Esteban:2020cvm, Gonzalez-Garcia:2021dve}. The atmospheric mixing angle $\sin^2{\theta_{23}}$ lies in the lower octant and marginally coincides with the lower bound of the experimental $3\sigma$ range\,\cite{Esteban:2020cvm, Gonzalez-Garcia:2021dve}. The Dirac CP phase $\delta$ is found to be $180^{\circ}$. It is to be highlighted that our model rules out the possibility of the inverted ordering of neutrino masses. Hence, in the light of normal ordering, the predicted upper bound on the sum of the neutrino masses\,($\sum{m_i}$, $i=1,2,3$) is consistent with the cosmological data\,\cite{Planck:2018vyg}. The upper bound on the three neutrino mass eigenvalues, $m_1$, $m_2$ and $m_3$ are $0.0265$ eV, $0.0279$ eV and $0.0575$ eV respectively. For this analysis, the input model parameters are tuned within certain numerical bounds: $0.0340\,\text{eV} <r < 0.0350\, \text{eV}$, $0.0270\,\text{eV}< r_3<0.0280\,\text{eV}$ and $-0.1125<x<-0.1120$. We show the maximum and minimum values of the predicted parameters in Table\,(\ref{table:3}). The graphical representations of the said predictions are shown in Figures \ref{fig:2(a)} - \ref{fig:2(g)}.
\begin{table}[h]
\centering
\begin{tabular}{c|c|c}
\hline
Predictions & Minimum Value & Maximum Value\\
\hline
$\sin^2 {\theta_{12}}$ &0.3402& 0.3414\\
\hline
$\sin^2 {\theta_{13}}$ & 0.0203 & 0.0238 \\
\hline
$\sin^2 {\theta_{23}}$ & 0.3900 & 0.4000 \\
\hline
$\delta$ & 180$^{\circ}$ & 180$^{\circ}$\\
\hline
$\Delta m^2_{21}$ &0.000035 $\text{eV}^2$ & 0.00011 $\text{eV}^2$ \\
\hline
$\Delta m^2_{31}$ & 0.00244 $\text{eV}^2$  & 0.00260 $\text{eV}^2$  \\
\hline
$m_1$ & 0.0257 eV & 0.0265 eV\\
\hline
$m_2$ & 0.0270 eV & 0.0279 eV\\
\hline
$m_3$ & 0.0558 eV & 0.0575 eV\\
\hline
$\sum{m_i}$ & 0.1085 eV & 0.1120 eV\\
\hline
\end{tabular}
\caption{Represents the approximate maximum and minimum values of the observable parameters} 
\label{table:3}
\end{table}

To illustrate the numerical values that the fundamental parameters of our model may adopt, we take the following example.

If we choose the cut-off scale $\Lambda$ to be approximately $4 \times 10^{13}$GeV, with other model parameters set as $v_\sigma = 10^{13}$ GeV, $v_\psi = 10^{12}$ GeV, $v_\phi = 250$ GeV, $v_h = 246$ GeV, $v_\kappa = 50$ GeV, $y_e = 0.2$, $y_\mu = 0.63$, $y_\tau = 0.66$, $|y_d| = 0.36$, $|y_s \pm y_a| = 0.045$, $\text{Arg}[y_d] = -23^\circ$, $\text{Arg}[y_s - y_a] = 37.81^\circ$, and $\text{Arg}[y_s + y_a] = 180^\circ$, the charged lepton masses are predicted to be $m_e = 0.51$ MeV, $m_\mu = 106$ MeV, and $m_\tau = 1777$ MeV, which are consistent with the experimental data\,\cite{ParticleDataGroup:2018ovx, ParticleDataGroup:2022pth}.

In addition, for the neutrino sector, these choices lead to the mass eigenvalues $m_1 = 0.0260$ eV, $m_2 = 0.0274$ eV, and $m_3 = 0.0564$ eV. The corresponding mass-squared differences are found approximately at their best-fit values, which are, $\Delta m^2_{21} = 0.0000742$ $\text{eV}^2$ \,\cite{Esteban:2020cvm, Gonzalez-Garcia:2021dve} and $\Delta m^2_{31} = 0.002515$ $\text{eV}^2$ \,\cite{Esteban:2020cvm, Gonzalez-Garcia:2021dve}.

The chosen VEVs can be achieved by tuning the twenty-five free parameters present in the five equations, namely, Eqs. (\ref{mu1}) - (\ref{mu5}), which are also highlighted in Appendix \ref{appB}. To illustrate this, we conduct a numerical scan, where we allow $\lambda$'s  and $\mu^2$ 's to vary randomly within the approximate ranges, $0.01 - 0.99$ and $4 \times 10^{11} \, \text{GeV}^2 - 1.2 \times 10^{26} \, \text{GeV}^2$, respectively.

\begin{figure}
  \centering
    \subfigure[]{\includegraphics[width=0.31
  \textwidth]{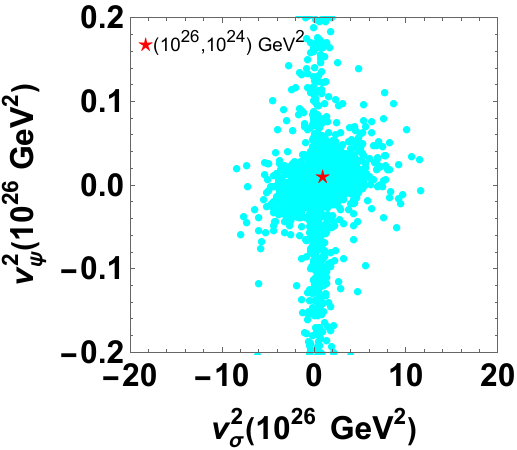}\label{fig:3(a)}} 
    \subfigure[]{\includegraphics[width=0.284
    \textwidth]{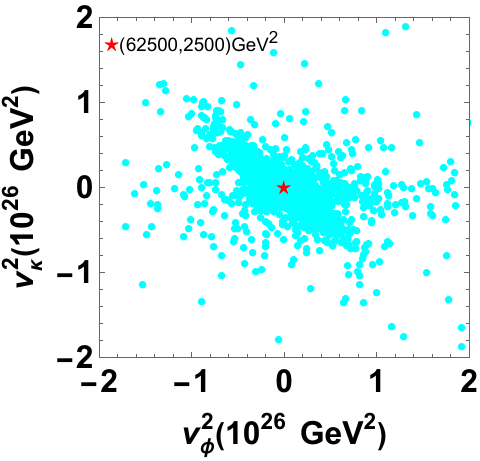}\label{fig:3(b)}} 
    \subfigure[]{\includegraphics[width=0.298
    \textwidth]{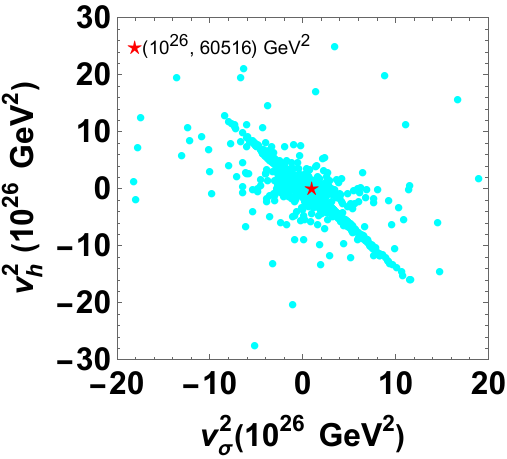}\label{fig:3(c)}} 
\caption{Plots showing the parameter space for $v^2$ 's. In the plots, `cyan dots' indicate the obtained parameter space, while `red stars' mark the﻿ chosen $v^2$ 's values.﻿}
\label{fig:3}
\end{figure}

For these randomly generated inputs, we obtain a large parameter space for the $v^2$ 's, as illustrated in the Figure \ref{fig:3}.
From the plots in Figure \ref{fig:3}, we observe that our choice of VEVs lie within the obtained parameter space. To emphasize this, the chosen VEVs\,($v^2$ 's) are highlighted with a `red star' in each plot of Figure \ref{fig:3}.

\section{Summary and Discussions \label{Summary}}
In summary, we have built a Dirac neutrino model. In this model, we have constructed a neutrino mass matrix in the framework of $\Delta(27)$ symmetry. We have enhanced the framework by additional cyclic symmetries, $Z_3$, $Z_7$ and $Z_{10}$.  The $Z_{10}$ and $Z_3$ symmetries help to restrict the unfavourable terms and NLO corrections from appearing in the Yukawa Lagrangian. In addition, the $Z_7$ symmetry plays a crucial role in forbidding possible Majorana mass terms in the model.

	A few salient features of our model are listed below:
\begin{itemize}
    \item The model contains a few flavons, and can explain the charged lepton mass hierarchy. It is predictive, with a single guiding parameter, $\rho$, dictating the three mixing angles and the Dirac CP phase, $\delta$.
    \item The model is consistent with the normal ordering of neutrino masses and it predicts the three neutrino mass eigenvalues, $m_1$, $m_2$ and $m_3$ to have upper bounds of $0.0265 $eV, $0.0279$eV, and $0.0575$ eV respectively. In addition, the upper bound on the sum of the three neutrino masses is found to be $0.112$eV, which is in good agreement with Planck's cosmological data\,\cite{Planck:2018vyg}.

    \item The model emphasizes on the \emph{standard parametrization} of the lepton mixing matrix,  which is often overlooked by model builders. This parametrization is important from a model-building point of view because a general unitary matrix contains nine parameters, out of which five are unphysical in the context of neutrino physics experiments. Therefore, we have taken special care to isolate these phases from our lepton mixing matrix to ensure an accurate mixing scheme.
\end{itemize}

Here we wish to highlight one important feature of the proposed model. In general, smallness of neutrino masses is attributed to the existence of a heavy fermion in the seesaw mechanism (in both Dirac or Majorana scenario). Here, we have tried to address the issue involving both (i) the smallness of neutrino mass and (ii) the hierarchy of charged leptons from a different perspective where both cut-off scales and the VEVs of the flavon fields finally fix the said issues. It is little bit disturbing at the first glance to see that $m_i \sim 1/ \Lambda^2$\,(where $i= 1, 2, 3$) and $m_l \sim 1/ \Lambda^7$\,(where $l= e, \mu, \tau$). However, one cannot ignore the involvement of VEVs, $v_\sigma^x v_{\phi}^y v_\psi ^{z}$ in the expression of both $m_i$ and $m_l$. The powers $x$, $y$ and $z$ appear differently for both cases and thus naturally lead to the desired results. Hence, $\Lambda$ should not be visualised as a suppression factor for the present discussion which might be true in other contexts. The involvement of cyclic groups strictly restricts any possibility of $1/\Lambda^n$, with $n>7$ to appear in theory. 

\vspace{0.2cm}
		
		The nature of neutrinos, whether they are Majorana or Dirac, remains uncertain. In this work, we have tried to show that, without assuming the Majorana nature of neutrinos or using the seesaw mechanism (either Majorana or Dirac), it is possible to explain the neutrino masses, mixing, mass hierarchies of neutrinos, and the charged lepton mass hierarchy by following the classical mass generation mechanism for fermions. 

\section{Acknowledgement} 
The authors thank R. Srivastava (IISER Bhopal), B. Karmakar (Institute of Physics, University of Silesia, Katowice), and D. Bhattacharjee and P. Chakraborty (Gauhati University) for fruitful discussions. The research work of MD is supported by the Council of Scientific and Industrial Research (CSIR), Government of India, through a NET Senior Research Fellowship under grant No. 09/0059(15346)/2022-EMR-I.

\section{Data availability statement}
The data cannot be made publicly available upon publication because no suitable repository exists for hosting data in this field of study. The data that support the findings of this study are available upon reasonable request from the authors.
%
\biboptions{sort&compress}
\bibliography{reference1.bib}
\section{Appendix A \label{appA}}
\subsection{$\Delta(27)$ group}
The multiplication rules under the $\Delta(27)$ group is given below,
\begin{eqnarray}
3 \otimes 3 &=& \bar{3}_{s_1} \oplus \bar{3}_{s_2} \oplus  \bar{3}_a, \nonumber\\
\bar{3} \otimes \bar{3} &=& 3_{s_1} \oplus  3_{s_2} \oplus  3_a, \nonumber\\
3 \otimes \bar{3} &=& \sum_{r=0}^2 1_{r,0} \oplus  \sum_{r=0}^2 1_{r,1} \oplus  \sum_{r=0}^2 1_{r,2},\nonumber\\
1_{p,q} \otimes 1_{p',q'} &=& 1_{(p+p')\,\text{mod}\,3,\, (q+q')\,\text{mod}\,3}.
\end{eqnarray}
If $(a_1, a_2, a_3)$ and $(b_1, b_2, b_3)$ are two triplets under $\Delta(27)$ then,
\begin{eqnarray}
(3 \otimes 3)_{\bar{3}_{s_1}} &=& (a_1 b_1 , a_2 b_2 , a_3 b_3),\nonumber\\
(3 \otimes 3)_{\bar{3}_{s_2}} &=& \frac{1}{2}(a_2 b_3 + a_3 b_2, a_3 b_1 + a_1 b_3, a_1 b_2 + a_2 b_1),\nonumber\\
(3 \otimes 3)_{\bar{3}_a} &=& \frac{1}{2}(a_2 b_3 - a_3 b_2, a_3 b_1 - a_1 b_3, a_1 b_2 - a_2 b_1),\nonumber\\
(3 \otimes \bar{3})_{1_{r,0}} &=& a_1 b_1 + \omega^{2r} a_2 b_2 + \omega^r a_3 b_3,\nonumber\\
(3 \otimes \bar{3})_{1_{r,1}} &=& a_1 b_2 + \omega^{2r} a_2 b_3 + \omega^r a_3 b_1, \nonumber\\
(3 \otimes \bar{3})_{1_{r,2}} &=& a_1 b_3 + \omega^{2r} a_2 b_1 + \omega^r a_3 b_2, 
\end{eqnarray}
where, $r= 0,1,2$ and $\omega= e^{2\pi i/3}$.
\subsection{$Z_{N}$ group}
$Z_{N}$ group, represents a symmetry that operates within the integers modulo N; it involves the numbers 0 to $(N-1)$.
Suppose, $n_1$ and $n_2$ are two group elements of $Z_{N}$, then,
\begin{equation}
n_1 \times n_2 = (n_1+n_2)\,mod\,N.
\end{equation}
The irreducible representation of a general group element\,($n$), is given by: $e^{2 n \pi i/N}$. In Tables~\ref{table:z3}, \ref{table:z7}, and \ref{table:z10}, we present the multiplication tables for the $Z_3$, $Z_7$, and $Z_{10}$ groups, respectively.

\begin{table}[!h]
\centering
\begin{tabular}{|p{0.35cm}|p{0.1cm}p{0.1cm}p{0.1cm}|}
\hline
$Z_{3}$& \,0 & 1 & 2  \\
 \hline
0& \,0 & 1 & 2 \\
\hline
1& \,1 & 2 & \textbf{0}  \\
\hline
2& \,2 & \textbf{0} & 1  \\
\hline
\end{tabular}
\caption{Represents the multiplication table of the $Z_{3} $ group.} 
\label{table:z3}
\end{table}

\begin{table}[!h]
\centering
\begin{tabular}{|p{0.35cm}|p{0.1cm}p{0.1cm}p{0.1cm}p{0.1cm}p{0.1cm}p{0.1cm}p{0.1cm}|}
\hline
$Z_{7}$& \,0 & 1 & 2 & 3 & 4 & 5 & 6 \\
 \hline
0& \,0 & 1 & 2 & 3 & 4 & 5 & 6 \\
\hline
1& \,1 & 2 & 3 & 4 & 5 & 6 & \textbf{0} \\
\hline
2& \,2 & 3 & 4 & 5 & 6 & \textbf{0} & 1 \\
\hline
3& \,3 & 4 & 5 & 6 & \textbf{0} & 1 & 2 \\
\hline
4& \,4 & 5 & 6 & \textbf{0} & 1 & 2 & 3 \\
\hline
5& \,5 & 6 & \textbf{0} & 1 & 2 & 3 & 4 \\
\hline
6& \,6 & \textbf{0} & 1 & 2 & 3 & 4 & 5 \\
\hline
\end{tabular}
\caption{Represents the multiplication table of the $Z_{7} $ group.} 
\label{table:z7}
\end{table}

\begin{table}[!h]
\centering
\begin{tabular}{|p{0.38cm}|p{0.1cm}p{0.1cm}p{0.1cm}p{0.1cm}p{0.1cm}p{0.1cm}p{0.1cm}p{0.1cm}p{0.1cm}p{0.1cm}|}
\hline
$Z_{10}$& \,0 & 1 & 2 & 3 & 4 & 5 & 6 & 7 & 8 & 9 \\
 \hline
0& \,0 & 1 & 2 & 3 & 4 & 5 & 6 & 7 & 8 & 9\\
\hline
1& \,1 & 2 & 3 & 4 & 5 & 6 & 7 & 8 & 9 & \textbf{0} \\
\hline
2& \,2 & 3 & 4 & 5 & 6 & 7 & 8 & 9 & \textbf{0} & 1 \\
\hline
3& \,3 & 4 & 5 & 6 & 7 & 8 & 9 & \textbf{0}& 1 & 2 \\
\hline
4& \,4 & 5 & 6 & 7 & 8 & 9 & \textbf{0} & 1 & 2 & 3 \\
\hline
5& \,5 & 6 & 7 & 8 & 9 & \textbf{0} & 1 & 2 & 3 & 4 \\
\hline
6& \,6 & 7 & 8 & 9 & \textbf{0} & 1 & 2 & 3 & 4 & 5 \\
\hline
7& \,7 & 8 & 9 & \textbf{0} & 1 & 2 & 3 & 4 & 5 & 6 \\
\hline
8& \,8 & 9 & \textbf{0} & 1 & 2 & 3 & 4 & 5 & 6 & 7 \\
\hline
9& \,9 & \textbf{0} & 1 & 2 & 3 & 4 & 5 & 6 & 7 & 8 \\
\hline
\end{tabular}
\caption{Represents the multiplication table of the $Z_{10} $ group.} 
\label{table:z10}
\end{table}
\section{Appendix B \label{appB}}
The scalar potential of the model, which is invariant under $\Delta(27)\times Z_{3} \times Z_7 \times Z_{10} \times SU(2)_L$ is presented below,
\begin{eqnarray}
V&=& V(H) + V(\psi)+ V(\phi)+ V(\kappa)+ V(\sigma)+ V(H,\psi)+ V(H,\phi)+ V(H,\kappa)+ V(H,\sigma)+ V(\psi, \phi)\nonumber\\&&+ V(\psi, \kappa)+ V(\psi, \sigma)+ V(\phi, \kappa)+ V(\phi, \sigma)+ V(\kappa, \sigma).
\end{eqnarray}
Writing the terms explicitly, we have,
\begin{align}
V(H)&=- \mu^2_H (H^{\dagger}H)+ \lambda^H_1 (H^{\dagger}H)^2,\nonumber\\
V(\psi)&=-\mu^2_{\psi} (\psi^{\dagger}\psi)+ \lambda^{\psi}_1 (\psi^{\dagger}\psi)_{1_{00}}(\psi^{\dagger}\psi)_{1_{00}}+\lambda^{\psi}_2 (\psi^{\dagger}\psi)_{1_{10}}(\psi^{\dagger}\psi)_{1_{20}}+\lambda^{\psi}_3 (\psi^{\dagger}\psi)_{1_{01}}(\psi^{\dagger}\psi)_{1_{02}}+ \lambda^{\psi}_4 (\psi^{\dagger}\psi)_{1_{11}}\nonumber\\&\quad\,(\psi^{\dagger}\psi)_{1_{22}}+\lambda^{\psi}_5 (\psi^{\dagger} \psi)_{1_{21}}(\psi^{\dagger}\psi)_{1_{12}},\nonumber\\
V(\phi)&=V(\psi \rightarrow \phi),\nonumber\\
V(\kappa)&=V(\psi \rightarrow \kappa),\nonumber\\
V(\sigma)&= V(H \rightarrow \sigma),\nonumber\\
V(H, \psi)&=\lambda^{H \psi} (H^{\dagger} H)(\psi^{\dagger}\psi),\nonumber\\
V(H, \phi)&=V(H, \psi \rightarrow H, \phi),\nonumber\\
V(H, \kappa)&=V(H, \psi \rightarrow H, \kappa),\nonumber\\
V(H, \sigma)&=\lambda^{H \sigma} (H^{\dagger} H)(\sigma^{\dagger}\sigma),\nonumber\\
V(\psi, \phi)&= \lambda^{\psi \phi}_1 (\psi^{\dagger}\psi)_{1_{00}}(\phi^{\dagger}\phi)_{1_{00}}+\lambda^{\psi \phi}_2 (\psi^{\dagger}\psi)_{1_{10}}(\phi^{\dagger}\phi)_{1_{20}}+\lambda^{\psi \phi}_3 (\psi^{\dagger}\psi)_{1_{01}}(\phi^{\dagger}\phi)_{1_{02}}+\lambda^{\psi \phi}_4 (\psi^{\dagger}\psi)_{1_{11}}(\phi^{\dagger}\phi)_{1_{22}}\nonumber\\&\quad\,+\lambda^{\psi \phi}_5 (\psi^{\dagger}\psi)_{1_{21}}(\phi^{\dagger}\phi)_{1_{12}}+\lambda^{\psi \phi}_6 (\psi^{\dagger}\psi)_{1_{20}}(\phi^{\dagger}\phi)_{1_{10}}+\lambda^{\psi \phi}_7 (\psi^{\dagger}\psi)_{1_{02}}(\phi^{\dagger}\phi)_{1_{01}}+\lambda^{\psi \phi}_8(\psi^{\dagger}\psi)_{1_{22}}(\phi^{\dagger}\phi)_{1_{11}}\nonumber\\&\quad\,+\lambda^{\psi \phi}_9 (\psi^{\dagger}\psi)_{1_{12}}(\phi^{\dagger}\phi)_{1_{21}},\nonumber\\
V(\psi, \kappa)&= V(\psi, \phi \rightarrow \psi, \kappa),\nonumber\\
V(\psi, \sigma)&= \lambda^{\psi \sigma} (\psi^{\dagger} \psi)(\sigma^{\dagger}\sigma),\nonumber\\
V(\phi, \kappa)&= V(\psi, \phi \rightarrow \phi, \kappa),\nonumber\\
V(\phi, \sigma)&= V(\psi, \sigma\rightarrow \phi, \sigma),\nonumber\\
V(\kappa, \sigma)&= V(\psi, \sigma \rightarrow \kappa, \sigma).\nonumber\\
\end{align} 
In models with discrete flavour symmetries, it is typical to have multiple coupling constants within the scalar potential. This flexibility allows for the selection of appropriate vacuum alignments for the scalar fields. For the sake of simplicity, we choose $\lambda _2^{\psi \phi }=\lambda _6^{\psi \phi }$ and $\lambda _2^{\phi \kappa }=\lambda _6^{\phi \kappa }$. We assume that the scalar fields develop the VEVs in the mentioned directions,  $\langle \psi \rangle = v_{\psi}(1,1,1)$, $\langle \phi \rangle = v_{\phi}(0,1,0)$, $\langle \kappa \rangle = v_{\kappa}(1,1,1)$, $\langle H \rangle = v_h$  and  $\langle \sigma \rangle = v_{\sigma}$, and the following minimisation equations are true,
\begin{align}
\frac{\partial V}{\partial H}&= v_h \left(2 v_h^2 \lambda _1^H-\mu _H^2+3 \lambda^{H \kappa} v_{\kappa }^2+3 \lambda ^{H \psi} v_{\psi }^2+\lambda ^{H \phi} v_{\phi }^2+ \lambda ^{H \sigma} v_{\sigma}^2\right)=0,\nonumber\\
\frac{\partial V}{\partial \psi_1}&=\frac{\partial V}{\partial \psi_3}=v_{\psi } \left(v_h^2 \lambda ^{H \psi}-\mu _{\psi }^2+3 \left(\lambda _1^{\psi \kappa }+\lambda _3^{\psi \kappa }+\lambda _7^{\psi \kappa }\right) v_{\kappa }^2+6 \left(\lambda _1^{\psi }+\lambda _3^{\psi }\right) v_{\psi }^2+\left(\lambda _1^{\psi \phi }-\lambda _2^{\psi \phi }\right) v_{\phi }^2+ \lambda ^{\psi \sigma } v_{\sigma }^2\right)=0,\nonumber\\
\frac{\partial V}{\partial \psi_2}&=v_{\psi } \left(v_h^2 \lambda ^{H \psi }-\mu _{\psi }^2+3 \left(\lambda _1^{\psi \kappa }+\lambda _3^{\psi \kappa }+\lambda _7^{\psi \kappa }\right) v_{\kappa }^2+6 \left(\lambda _1^{\psi }+\lambda _3^{\psi }\right) v_{\psi }^2+\left(\lambda _1^{\psi \phi }+2 \lambda _2^{\psi \phi }\right) v_{\phi }^2 +\lambda ^{\psi \sigma } v_{\sigma }^2\right)=0,\nonumber\\
\frac{\partial V}{\partial \phi_1}&=\frac{\partial V}{\partial \phi_3}=3 v_{\phi } \left(\lambda _7^{\phi \kappa } v_{\kappa }^2+\lambda _3^{\psi \phi } v_{\psi }^2\right)=0,\nonumber\\
\frac{\partial V}{\partial \phi_2}&=v_{\phi } \left(v_h^2 \lambda ^{H \phi}-\mu _{\phi }^2+3 \lambda _1^{\phi \kappa } v_{\kappa }^2+3 \lambda _1^{\psi \phi } v_{\psi }^2+2 \left(\lambda _1^{\phi }+\lambda _2^{\phi }\right) v_{\phi }^2+ \lambda ^{\phi \sigma } v_{\sigma}^2\right)=0,\nonumber\\
\frac{\partial V}{\partial \kappa_1}&=\frac{\partial V}{\partial \kappa_3}=v_{\kappa } \left(v_h^2 \lambda ^{H \kappa}-\mu _{\kappa }^2+6 \left(\lambda _1^{\kappa }+\lambda _3^{\kappa }\right) v_{\kappa }^2+3 \left(\lambda _1^{\psi \kappa }+\lambda _3^{\psi \kappa }+\lambda _7^{\psi \kappa }\right) v_{\psi }^2+\left(\lambda _1^{\phi \kappa }-\lambda _2^{\phi \kappa }\right) v_{\phi }^2+\lambda ^{\kappa \sigma} v_{\sigma}^2\right)=0,\nonumber\\
\frac{\partial V}{\partial \kappa_2}&=v_{\kappa } \left(v_h^2 \lambda ^{H \kappa}-\mu _{\kappa }^2+6 \left(\lambda _1^{\kappa }+\lambda _3^{\kappa }\right) v_{\kappa }^2+3 \left(\lambda _1^{\psi \kappa }+\lambda _3^{\psi \kappa }+\lambda _7^{\psi \kappa }\right) v_{\psi }^2+\left(\lambda _1^{\phi \kappa }+2 \lambda _2^{\phi \kappa }\right) v_{\phi }^2+\lambda ^{\kappa \sigma} v_{\sigma}^2\right)=0,\nonumber\\
\frac{\partial V}{\partial \sigma}&=v_{\sigma} \left(v_h^2 \lambda ^{H \sigma}+3 v_{\kappa}^2 \lambda ^{\kappa \sigma}-\mu_{\sigma} ^2+\lambda ^{\phi \sigma } v_{\phi}^2+3 \lambda ^{\psi \sigma } v_{\psi}^2+2 \lambda _1^{\sigma} v_{\sigma}^2\right)=0.
\end{align}
We solve the equations further, and observe that, among the many sets of solutions which are possible, we choose a set that is consistent with our framework. Therefore, for $\lambda^{\psi \phi}_2=\lambda^{\psi \phi}_3=\lambda^{\phi\kappa}_7=\lambda^{\phi \kappa}_2=0$, the following expressions are true,
\begin{align}
\mu_H^2&= 2 \lambda^H_1 v_h^2+3 \lambda ^{H \kappa} v_{\kappa}^2+\lambda ^{H \sigma} v_{\sigma}^2+3 \lambda ^{H \psi} v_{\psi}^2+\lambda ^{H \phi} v_{\phi}^2,\nonumber\\
\mu^2_{\psi}&=6( \lambda _1^{\psi }+ \lambda _3^{\psi })v^2_{\psi}+v_h^2 \lambda ^{H\psi}+3 v_{\kappa}^2( \lambda _1^{\psi \kappa }+ \lambda _3^{\psi \kappa }+\lambda _7^{\psi \kappa })+\lambda _1^{\psi \phi } v_{\phi}^2+\lambda ^{\psi \sigma } v_{\sigma}^2,\nonumber\\
\mu _{\phi }^2&=2(\lambda _1^{\phi }+\lambda _2^{\phi })v^2_{\phi}+v_h^2 \lambda ^{H\phi}+3 v_{\kappa}^2 \lambda _1^{\phi \kappa }+3 \lambda _1^{\psi \phi } v_{\psi}^2+\lambda ^{\phi \sigma } v_{\sigma}^2,\nonumber\\
\mu _{\kappa }^2&=6(\lambda _1^{\kappa }+\lambda _3^{\kappa })v^2_{\kappa}+v_h^2 \lambda ^{H\kappa}+\lambda ^{\kappa \sigma} v_{\sigma}^2+\lambda _1^{\phi \kappa } v_{\phi}^2+3 v_{\psi}^2(\lambda _1^{\psi \kappa } +\lambda _3^{\psi \kappa }+ \lambda _7^{\psi \kappa }),\nonumber\\
\mu_{\sigma}^2&=2 \lambda^{\sigma}_1 v^2_{\sigma}+v_h^2 \lambda ^{H \sigma}+3 v_{\kappa}^2 \lambda^{\kappa \sigma}+\lambda^{\phi \sigma } v_{\phi}^2+3 \lambda ^{\psi \sigma } v_{\psi}^2.\nonumber
\end{align}
The above equations are also mentioned in the text from Eqs.\,(\ref{mu1})\,- (\ref{mu5}).
\end{document}